\documentclass[review,onecolumn,12pt,a4paper,number]{elsarticle}

\usepackage{amsmath}
\usepackage{amssymb}
\usepackage{graphicx}
\usepackage{textcomp}
\usepackage{gensymb}
\usepackage[tight,nice]{units}
\usepackage{bm}
\usepackage{lineno}

\usepackage[caption=false]{subfig}
\usepackage[english]{babel}

\let\olditemize=\itemize
\def\itemize{
\olditemize \setlength{\itemsep}{-0.5ex} }
\let\oldenumerate=\enumerate
\def\enumerate{
\oldenumerate \setlength{\itemsep}{-0.5ex} }

\newcommand{\mum}{\ensuremath{\micro \mathrm{m} }}

\journal{Nuclear Instruments and Methods in Physics Research Section A}

\begin{document}

\begin{frontmatter}

\title{Copper coated carbon fiber reinforced plastics for high and ultra high
vacuum applications}

\author[PSI]{F.~Burri}
\author[PSI]{M.~Fertl}
\author[Galvanic]{P.~Feusi}
\author[PSI]{R.~Henneck}
\author[PSI,ETH]{K.~Kirch}
\author[PSI]{B.~Lauss}
\author[PSI]{P.~R\"{u}ttimann}
\author[PSI]{P.~Schmidt-Wellenburg\corref{cor1}}
\ead{philipp.schmidt-wellenburg@psi.ch}
\author[PTB]{A.~Schnabel}
\author[PTB]{J.~Voigt}
\author[PSI,IKC]{J.~Zenner}
\author[PSI]{G.~Zsigmond}

\cortext[cor1]{Corresponding author.}

\address[PSI]{Paul Scherrer Institut, 5232 Villigen, Switzerland}
\address[Galvanic]{Galvanic W\"{a}denswil Feusi\,+\,Federer AG, Zugerstrasse 160, 8820 W\"{a}denswil, Switzerland}
\address[ETH]{ETH Z\"{u}rich, Z\"{u}rich, Switzerland}
\address[PTB]{Physikalisch-Technische Bundesanstalt, 10587 Berlin, Germany}
\address[IKC]{Institute for Nuclear Chemistry, University of Mainz, Mainz, Germany}

\begin{abstract}We have used copper-coated carbon fiber reinforced plastic (CuCFRP) for the construction of high and
ultra-high vacuum recipients. The vacuum performance is found to be comparable to typical stainless steel
used for this purpose. In test recipients we have reached pressures of $\unit[2\!\times\!10^{-8}]{mbar}$ and measured a desorption
rate of $\unit[1\!\times\!10^{-11}]{mbar\!\cdot\!liter/s/cm^2}$; no degradation over time (2 years) has been found. Suitability for baking has been found to depend on the CFRP production process, presumably on the temperature of the autoclave curing. Together with other
unique properties of CuCFRP such as low weight and being nearly non-magnetic, this makes it an ideal
material for many high-end vacuum applications. 
\end{abstract}

\begin{keyword}
Vacuum recipient, hybrid coating, carbon fiber reinforced plastic, copper coating
\end{keyword}

\end{frontmatter}

\section*{Introduction}
Many of today's technologies and research techniques depend on processes and methods in high or ultrahigh vacuum.
Traditionally vacuum recipients are most often made of a variety of metals especially of various grades of stainless steel.
However, in certain applications stainless steel can be problematic as it is magnetic or at least magnetizable, has a relatively high weight, and is electrically conductive.

At the Paul Scherrer Institut, Switzerland an experiment to search for a permanent electric dipole moment of the neutron~(nEDM) is running\,\cite{Baker2011}, while ten time more sensitive next generation experiment is being setup.
The measurements are performed inside a magnetically shielded room in vacuum and are extremely sensitive to any perturbation or inhomogeneity of the magnetic field. 
It is of paramount importance to avoid all magnetic or magnetizable materials within the vacuum chamber.
Even parts made of non-magnetic stainless steel or aluminum are disfavored as electrical conductivity leads to Johnson-Nyquist noise\,\cite{Varpula1984}.
The spectral density of Johnson-Nyquist noise is proportional to
the thickness of a conductive layer, hence we studied vacuum and magnetic properties of micrometer thin
copper coatings on high performance carbon fiber reinforced plastic~(CFRP).

\section*{Requirements and samples}
Any material of and within the vacuum chamber has to fulfill among others the following stringent requirements:

\begin{itemize}
	\item No measurable magnetic impurities and magnetizability (local magnetic moments $M\!<\!\unit[5]{nAm^2}$ before and after contact with strong magnet).
	\item Low magnetic spectral noise density ($\leq \unit[20]{fT/\sqrt{Hz}}$)
	\item Very low bulk electric conductivity ($\sigma\!<\!\unit[10^4]{S/m}$),
or a very thin conductive layer on insulator ($d\!<\!\unit[20]{\mum}$).
	\item No helium permeability.
	\item Desorption rate below $\unit[10^{-10}]{mbar\!\cdot\!liter/s/cm^2}$.
\end{itemize}

\begin{figure}%
	\subfloat[CFRP sample before coating (front and backside).\label{fig:1a}]{\includegraphics[width=0.9\linewidth]{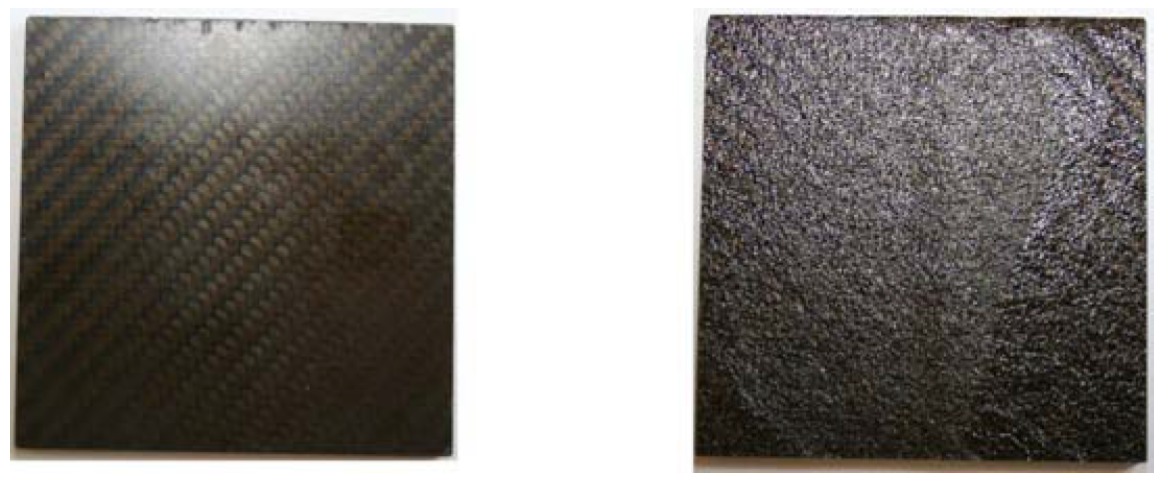}}\\
	\subfloat[Cu coated CFRP vacuum recipient.\label{fig:1b}]{\includegraphics[width=0.9\linewidth]{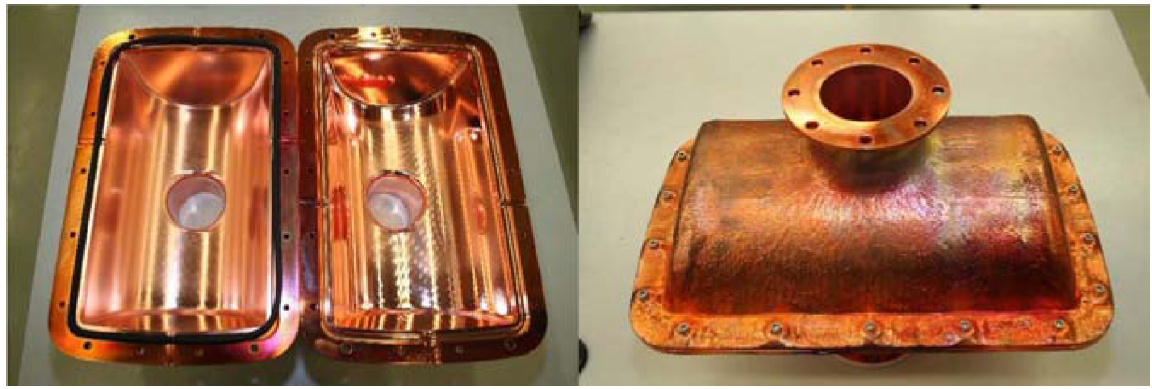}}\\
	\subfloat[Cu coated CFRP vacuum tube.\label{fig:1c}]{\includegraphics[width=0.9\linewidth]{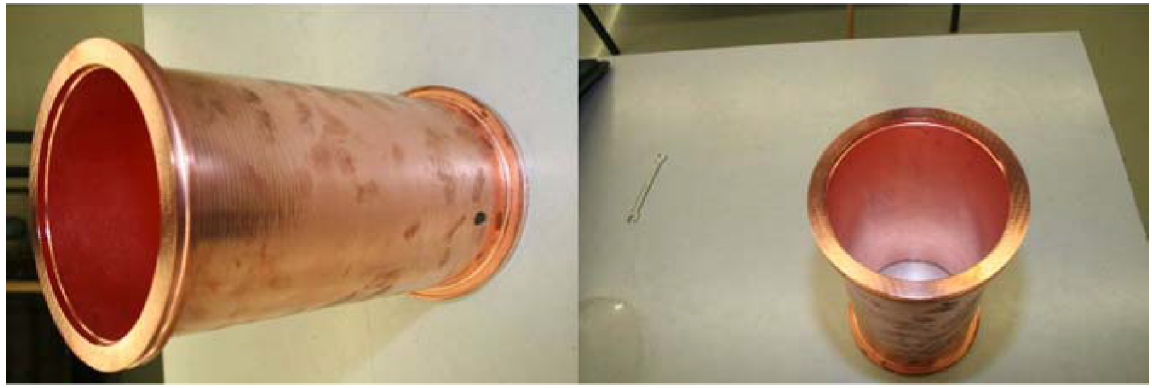}}
	\caption{Three different test pieces were used to investigate the vacuum properties of copper-coated CFRP\@. The square CFRP sample~(a) was first characterized without coating, later with. A custom made model~(b) of a possible later application was made using one aluminum negative to make two half shells. A standard, ``off-the-shelf'', CRFP tube~(c) was fitted with
two custom made ISO\,K\,DN160 flanges to demonstrate that the technology can be used to produce standard vacuum parts.}%
	\label{fig:1}%
\end{figure}

Carbon fiber reinforced plastics are known to have excellent mechanical properties and can be tailored to design.
This allows to reduce weight and still excel in strength of shape. Three different samples (Fig.\,\ref{fig:1}) were produced to investigate the properties of copper-coated CFRP: A quadratic plate ($\unit[100\!\times\!100\!\times\!5]{mm^3}$), a custom designed and tailored vacuum recipient made from high performance CFRP, and a standard vacuum tube made from lower grade CFRP with inside diameter~\unit[150]{mm}\footnote{Produced by C-Tech, PO Box 71-131, Rosebank, 1348, Auckland, NEW ZEALAND} with
two ISOK DN160 flange\footnote{All custom made parts were produced by Pauco Plast http://www.paucoplast.ch/, using LTM12/CF 0300 LTM12/CF 0700.}\@.

\begin{figure}%
	\includegraphics[width=0.9\columnwidth]{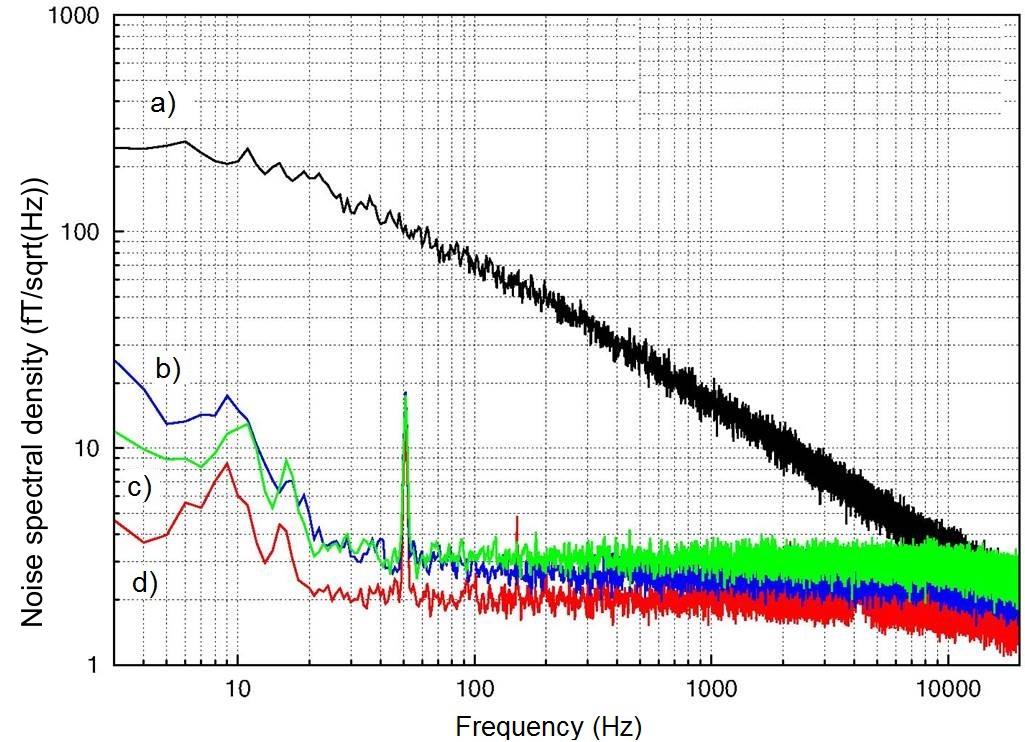}%
\caption{Comparison of spectral noise measurements using an array of SQUIDS at PTB\,\cite{Burghoff2007}. The noise spectrum of the actually used Al electrode (thickness \unit[30]{mm}~(a) was an order of magnitude larger than of both CFRP samples: b)~uncoated CFRP, c)~\unit[500]{nm} Cu on CFRP\@. The system noise~(d) is shown as reference. Note, that this CuCFRP sample was produced using magnetron sputtering; galvanically coated CFRP typically has a Cu-layer thickness of $\sim \unit[20]{\mum}$ and hence a slightly higher noise spectrum.}%
\label{fig:2}%
\end{figure}

The suitability of the material with respect to its magnetic properties was shown in the BMSR-II of the PTB in Berlin\,\cite{Bork2000}. Figure\,\ref{fig:2} compares the noise spectrum of a CFRP plate, a copper-coated CFRP plate, and an aluminum electrode. 
Both CFRP samples showed no dipole structure and a peak to peak magnetic field of $B<\unit[40]{pT}$ in a distance of $\sim \unit[3]{cm}$ even after magnetization with a strong permanent magnet. We consider the material suitable for our magnetic requirements and will concentrate on the vacuum properties in this article.

\section*{Sample characterization}
\begin{figure}%
	\includegraphics[width=0.9\columnwidth]{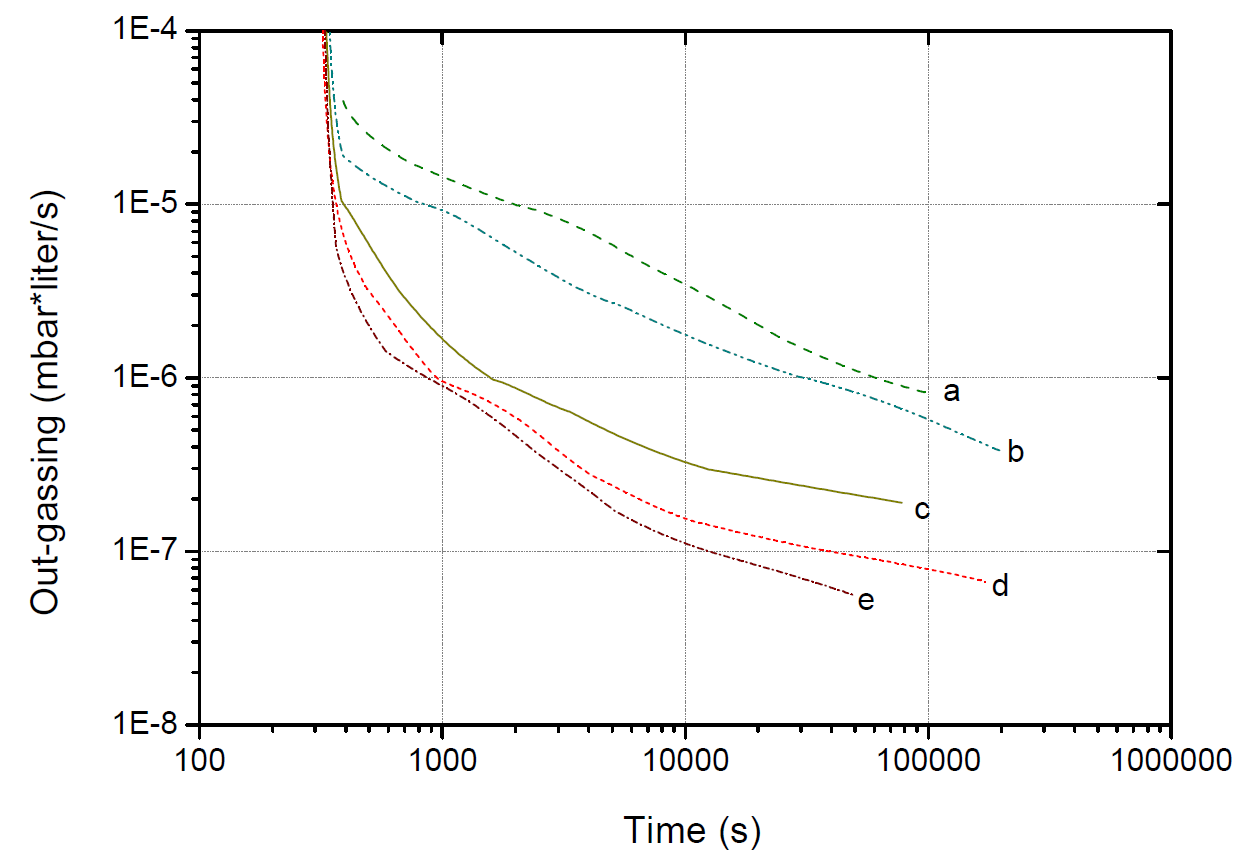}%
\caption{Out-gassing of a $\unit[10\!\times\!10]{cm^2}$ CFRP sample in a test vacuum recipient. a) First vacuum pumping after cleaning in purified benzine; b) second vacuum pumping after leaving sample in air for \unit[24]{hours}; c) after baking for \unit[24]{hours} at \unit[100]{\celsius} and exposing to air for \unit[1]{min}. The last two curves have been measured after coating the sample galvanically with \unit[20]{\mum} of copper: d) First
vacuum pumping after coating, e) second after baking the sample at \unit[100]{\celsius} and exposing it to air for \unit[2]{hours}.}%
\label{fig:3}%
\end{figure}

\begin{figure}%
	\includegraphics[width=0.9\columnwidth]{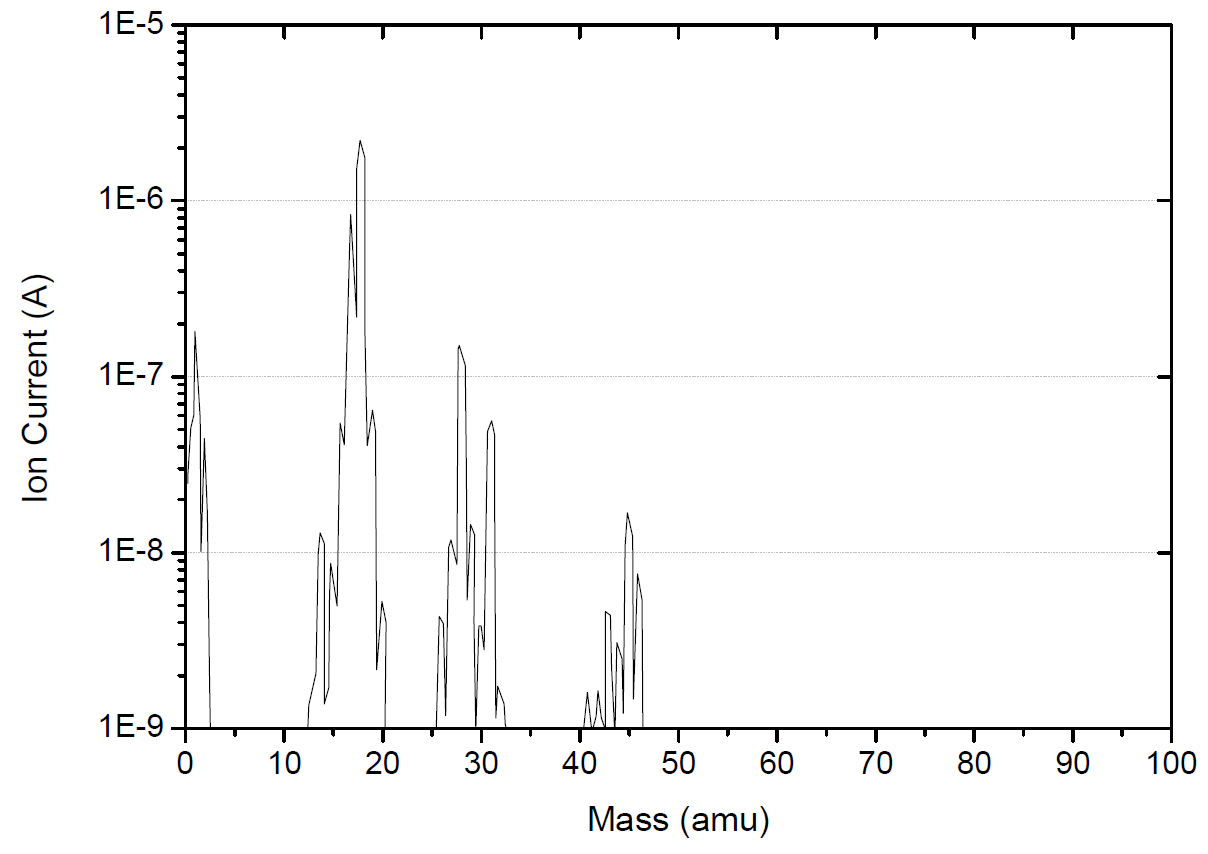}%
\caption{Residual gas composition of CFRP sample measured with a Pfeiffer PRISMA quadrupole mass spectrometer. More than \unit[90]{\%} is water, all other peaks are typical for rest gas. Note that no masses above ethanol~(46) were found.}%
\label{fig:4}%
\end{figure}

In an initial step we determined the out-gassing characteristics of the sample plate without coating. 
The plate was cleaned in an ultrasonic bath of pure benzine. 
Then the out-gassing rate was measured in a vacuum recipient, connected to a turbo pump (TPH 330) and roughing pump (Leybold SC5) with a pump speed of \unit[200]{liter/s}.
The pressure, which in our geometry using a $\unit[10\!\times\!10]{cm^2}$ ($\sim \unit[200]{cm^2}$ surface) sample easily
relates to the out-gassing rate, was measured using an ionization gauge (TPG\,300/IKR\,020, Balzers). Figure\,\ref{fig:3} shows three measured curves for the uncoated CFRP plate: (a)~sample directly after cleaning, (b)~after a first vacuum pump down and subsequent exposure to air for 24 h, and (c)~after baking for \unit[24]{h} and \unit[1]{min} exposure to air. We note here that baking improved the surface properties of all Cu coated pieces considerably making them more resistant, e.g.\ against finger prints.
We also characterized the composition of the rest gas in case (a) using a quadrupole mass spectrometer PRISMA\footnote{Pfeiffer Vacuum Technology}(see Fig.\,\ref{fig:4}), which is dominated by over \unit[90]{\%} of water.
Next the sample was coated by Galvanic W\"{a}denswil\footnote{http://www.galvanic.ch/} using a proprietary galvanic procedure.
This unique technique avoids any aggressive or implantation preparation processes, guaranteeing the conservation of the surface quality and maintaining the non-magnetic properties of the sample.
The coating decreased the out-gassing by a factor of ten which further improved by a factor of two after baking the sample.
We also studied the helium permeability in a dedicated setup.
For this purpose the CuCFRP plate was clamped directly onto a flange of a vacuum leak detector using a fluoroelastomer sealing.
The flange, clamps, and plate were then put into a tightly closed plastic bag filled with helium.
The integral leakage rate did not increase from the regular background rate of the leak detector when blind flanged with a fluoroelastomer sealing ($\unit[1\!\times\!10^{-7}]{mbar\!\cdot\!liter/s}$).
Hence, no permeability was detected for the CuCFRP plate, instead the leakage rate of the test assembly was identical to that of a fluoroelastomer sealed flange.

\section*{Vacuum recipient}
\begin{figure}%
	\includegraphics[width=0.9\columnwidth]{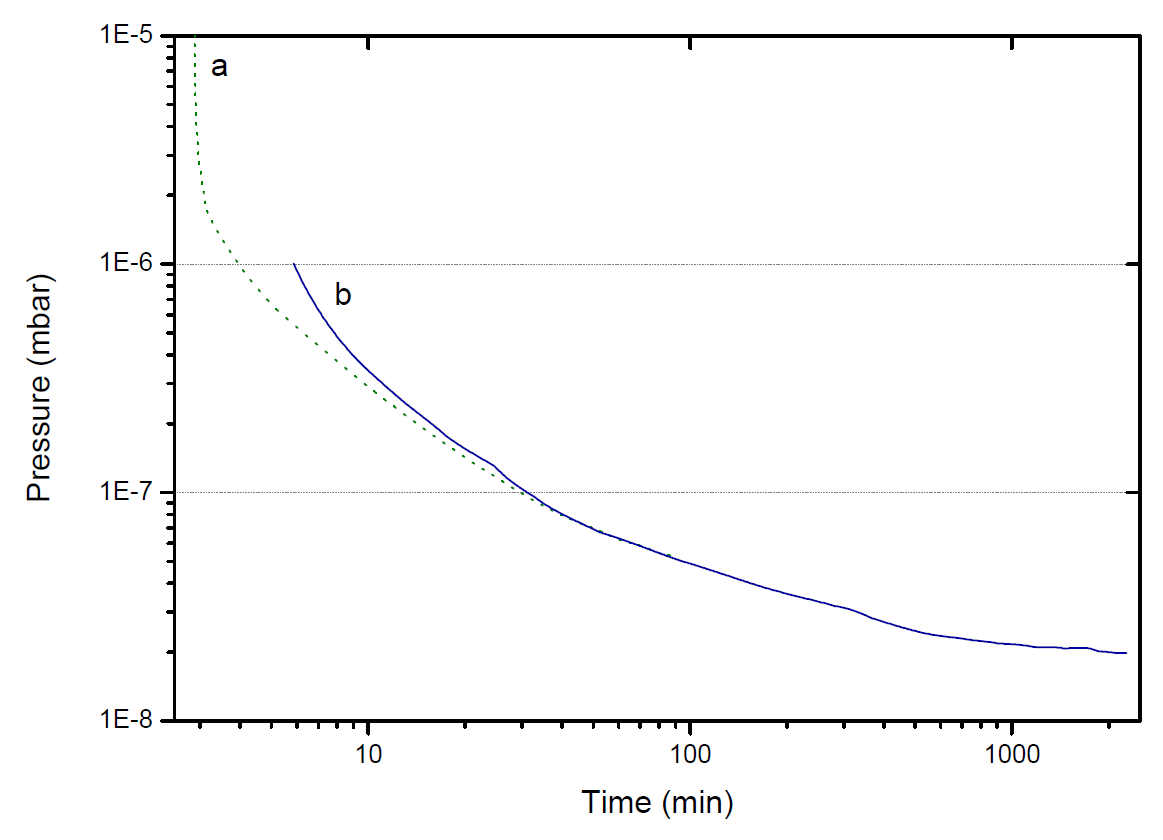}%
\caption{Evacuation curve of the custom made CuCFRP vacuum recipient (Fig\,\ref{fig:1b}). (a) After baking a pressure of $\unit[5\!\times\!10^{-8}]{mbar}$ was reached after \unit[1.5]{hours}; (b) after \unit[2]{hours} venting a second evacuation resulted in a nearly identical curve with a final
pressure of $\unit[2\!\times\!10^{-8}]{mbar}$ reached after one day of pumping.}%
\label{fig:5}%
\end{figure}

The second test piece was a custom design of a scaled down model of a possible vacuum recipient for our experiment, shown in Fig.\,\ref{fig:1b}. The size ($\unit[420\!\times\!270\!\times\!300]{mm^3}$) was chosen such that the half shells could fit
into available galvanization test baths.
On each half shell an ISO\,K DN63 
flange was attached to demonstrate that a connectivity to standard vacuum parts can be accomplished. The recipient was produced
by Pauco Plast\footnotemark[2] in an autoclave at \unit[200]{\celsius} using a negative mold made from aluminum. Both half shells were
then coated by Galvanic W\"{a}denswil\footnotemark[4]. The recipient made of these two half shells was tested in the same way for permeability as the small sample plate. Again, the results indicated that only the fluoroelastomar seals contributed to the leakage rate.
The recipient was then connected directly to a turbo molecular pump (TMH260, Pfeiffer) backed by a membrane pump (pumping speed $\unit[1]{m^3/s}$, end pressure $\sim \unit[1]{mbar}$). It was baked at \unit[100]{\celsius} for \unit[18]{hours} during evacuation.
After this time the pressure measured on the second flange of the recipient was $\unit[1.7\!\times\!10^{-7}]{mbar}$; a typical value for such a pump configuration.
In a next step we increased the temperature further to \unit[130]{\celsius} for \unit[6]{hours} before we let it cool down to room temperature. The mass spectrum of the rest gas was dominated
by water (\unit[95]{\%}) and nitrogen (\unit[5]{\%}). Carbohydrates and other heavy molecules made up less than \unit[0.1]{\%}.
The total pressure after cool down was identical to pressures reached by identical pumps blind 
flanged using fluoroelastomer seals. Figure\,\ref{fig:5} shows two subsequent evacuation curves. A final pressures of $\unit[5\!\times\!10^{-8}]{mbar}$ was reached after baking and \unit[100]{min} of pumping. A second pumping curve, recorded after venting and exposure to air for \unit[2]{hour}, coincides with the first curve after \unit[90]{min}. The final pressure reached was $\unit[2\!\times\!10^{-8}]{mbar}$.

\section*{Vacuum tube}
\begin{figure}%
	\includegraphics[width=0.9\columnwidth]{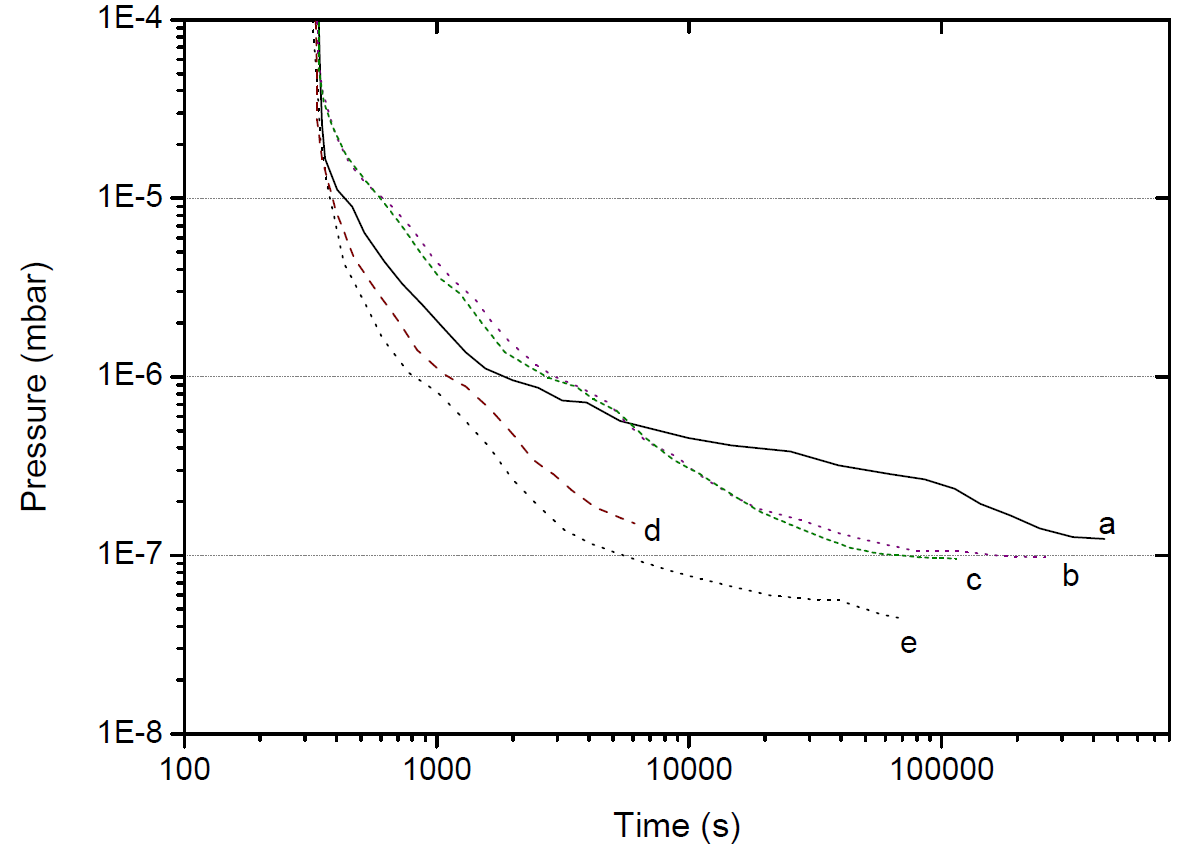}%
\caption{Evacuation curves of the tube made of copper-coated CFRP compared to a stainless steel tube~(c). Evacuation curve
without recipient~(e); with CuCFRP tube before cleaning~(a); CuCFRP tube cleaned with alcohol~(b); second evacuation of cleaned CuCFRP tube after 10 min venting~(d).}%
\label{fig:6}%
\end{figure}

The third test sample was an ``off-the-shelf'' CFRP tube\footnotemark[1] glued to two ISO\,K DN160 end 
flanges produced by Pauco Plast\footnotemark[2] entirely made of CFRP, and later copper-coated by Galvanic W\"{a}denswil\footnotemark[4], shown in Fig.\,\ref{fig:1c}. The tube was cleaned with a cotton cloth and purified ethanol before evacuation. Vacuum tightness and He-permeability of the ISO\,K\,DN160 flange sealed with an fluorelastomer seal was tested with a leak tester (HLT560, Pfeiffer). Only a very small permeability through the gasket of $<\!\unit[10^{-8}]{mbar\!\cdot\!liter/s}$ could be measured.
The tube was then attached with a locking collar to a turbo molecular pump (TPH 330, Pfeiffer)
backed by a Leybold SC\,15 scroll pump with a total pumping speed of \unit[200]{liter/s}. Several evacuation runs have been performed. For comparison an identical tube made from stainless steel~(316L) was also evacuated.
Figure\,\ref{fig:6} shows the pressure curves over time of these vacuum tests.
It was remarkable that after cleaning with ethanol the CuCFRP tube behaved similar to the standard stainless steel tube.

\begin{figure}%
	\includegraphics[width=0.9\columnwidth]{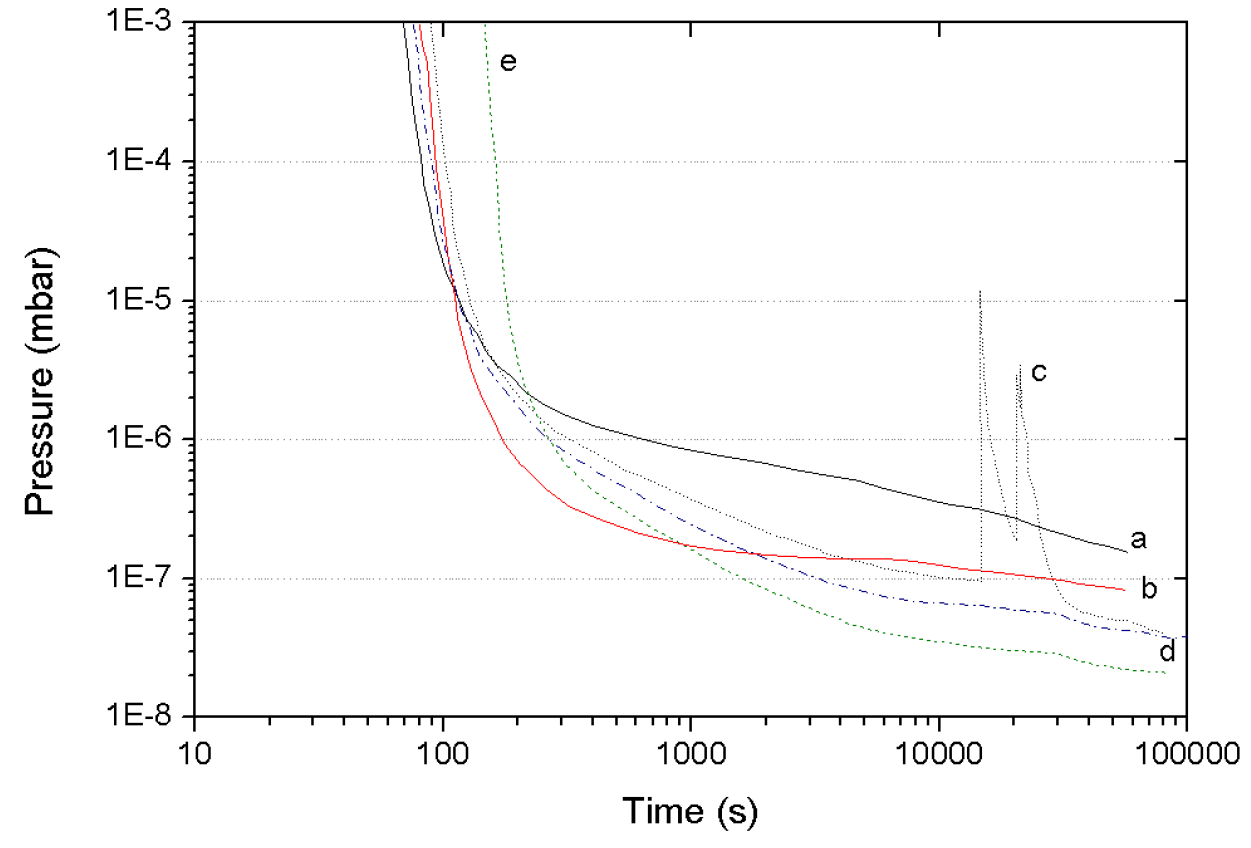}%
\caption{Evacuation curves of CuCFRP tube after two years storage in air. First evacuation of tube after storage~(a); second
evacuation after venting with $\text{N}_2$~(b); after baking the stainless steel
flanges (see Fig.\,\ref{fig:8}) with a hot air gun~(c); fourth evacuation after baking end flange and venting with air~(d); evacuation without CuCFRP tube only with flange and pressure gauge~(e).}%
\label{fig:7}%
\end{figure}

\subsection*{Performance stability of the tube}
With excellent performance of the custom made vacuum recipient (Fig.\,\ref{fig:1b}) in terms of long term stability and baking, we next investigated the long term stability of the coated CRFP tube.

Two years after initial characterization, storing the tube at air, we remeasured the evacuation behavior (see Fig.\,\ref{fig:7}) of the CuCFRP tube using a turbo molecular pump (TMH521, Pfeiffer) backed by a Leybold EcoDry\,$\unit[15]{m^3}$ with a pumping speed of \unit[500]{liter/s}. After baking the stainless steel end flanges a end pressure of less then $\unit[4\,\times\,10^{-8}]{mbar}$ was reached.
A desorption rate of about $\unit[10^{-11}]{mbar\!\cdot\!liter/s/cm^2}$ for the CuCRRP tube was deduced by subtracting the measured desorption rates of the pump alone and pump and tube.

Next we tested the sample durability with an `accelerated lifetime test' : {\it i)}~thermal cycling between $30$ and \unit[110]{\celsius} under vacuum, {\it ii)}~pressure cycling, i.e.\ the internal pressure was alternated between `zero' (i.e.\ vacuum) and atmospheric pressure in order to exert mechanical stress, especially on the connections between tube and end flanges.

\begin{figure}%
		\includegraphics[width=0.9\columnwidth]{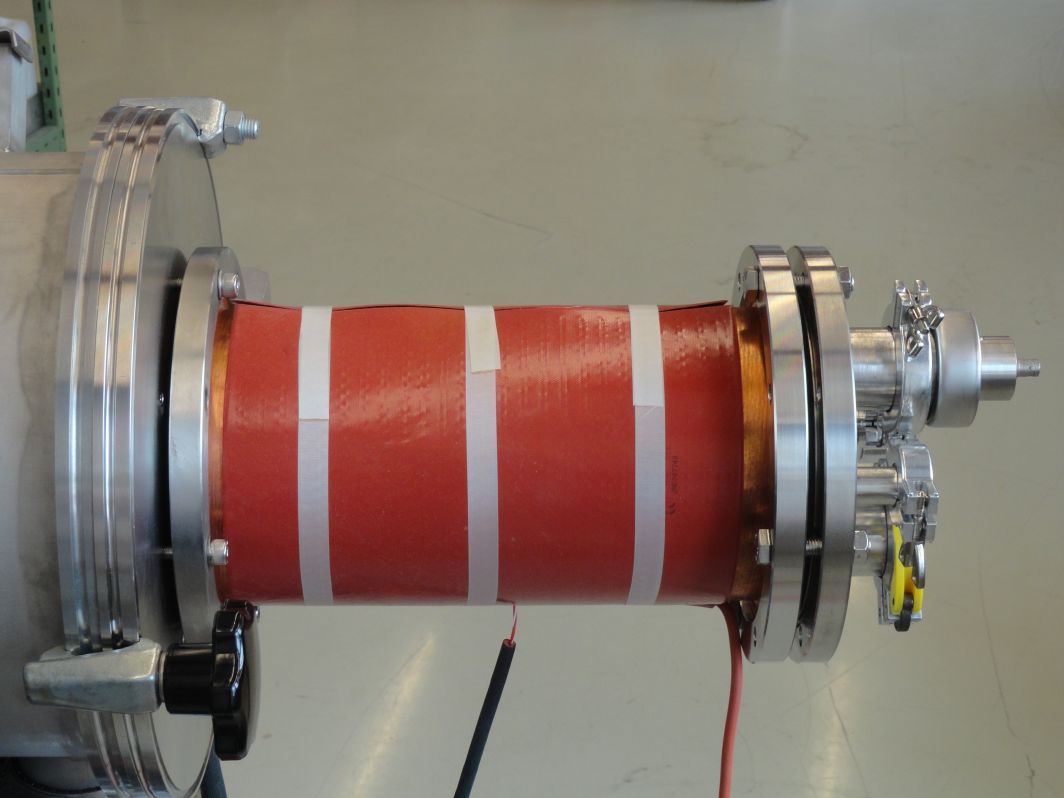}%
		\caption{Photograph of the setup with thermal insulation removed. One can see that the heating blanket does not extend all the way to the end flanges but leaves about $1$ to \unit[2]{cm} of open space.}%
		\label{fig:8}%
\end{figure}

{\it i)} The tube was flanged to a stainless steel vessel with a volume of \unit[40.2]{liter}
pumped by a turbo pump (Pfeiffer TMH~071\,Y\,P) and closed with a stainless steel end flange with a connected vacuum gauge. We estimated the pumping speed at the connection as about \unit[30]{liter/s}. As sealing we used fluoroelastomar O-rings. The tube (excluding the end flanges) was completely covered by a heatable blanket which could be controlled at a preset temperature via a PT~100 sensor. Figure\,\ref{fig:8} shows a picture of the setup. One can see that the heating blanket did not extend all the way to the end flanges but left about $1$ to \unit[2]{cm} of open space. Since both end flanges were not heated - and the overall thermal insulation enclosing the whole setup (not shown in Fig.\,\ref{fig:8}) was inefficient - this meant that we had considerable temperature gradients over the regions near the end flanges. Which in turn transformed into mechanical stress in these regions.

\begin{figure}%
		\includegraphics[width=\columnwidth]{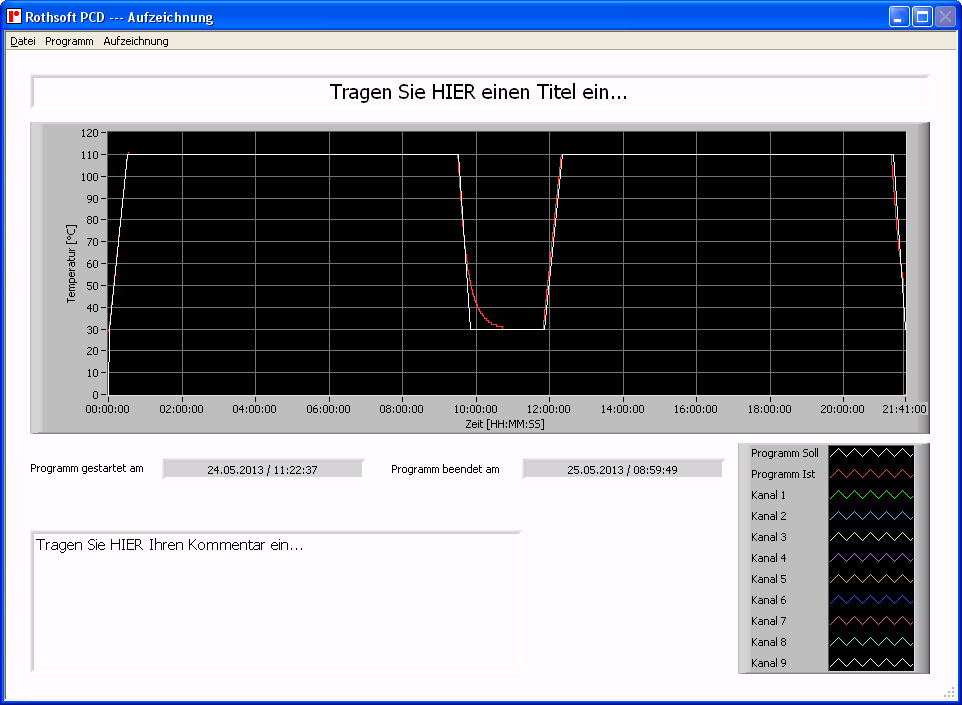}%
		\caption{Evolution of temperature (a)~set temperature, b)~measured temperature) and pressure~(c) as a function of time for the first two heating cycles. Due to the small heat capacity of the tube the temperature of the tube closely reflected the heating pattern while the internal pressure lagged behind by about \unit[30]{min}. The occasionally visible pressure `spikes' were interpreted as small internal leaks, when tiny cracks opened in the Cu-coating and a small amount of nitrogen from the previous flushing with nitrogen (as verified by a quadrupole mass spectrometer) was released.}%
		\label{fig:9}%
\end{figure}

Thermal cycling was started after two weeks of pumping at room temperature at an initial pressure of $\unit[2.9\!\times\!10^{-7}]{mbar}$. The tube was heated to \unit[110]{\celsius} within \unit[30]{min} and kept there for about \unit[10]{hours}.
Cooling to \unit[30]{\celsius} was effected within \unit[30]{min} and the lower temperature was kept then for \unit[2]{h} until the next cycle started. Figure\,\ref{fig:9} shows the evolution of pressure and temperature as a function of time. Due to the small heat capacity of the tube the temperature of the tube as well as the internal pressure closely reflected the heating pattern. A total of 23 cycles was performed. Two days after the last heating the pressure read $\unit[2.2\!\times\!10^{-7}]{mbar}$, slightly lower than at the beginning of the heat test. Afterwards, the test tube was disconnected and the stainless steel vessel was pumped alone to a marginal lower pressure of $\unit[2.0\!\times\!10^{-7}]{mbar}$.

{\it ii)} After the first seven temperature cycles and at the end of thermal cycling we performed the pressure cycling, all in all 25 cycles. The whole setup was repeatedly flushed with dry nitrogen to atmospheric pressure and then pumped to a pressure below \unit[$10^{-5}$]{mbar}.

\begin{figure}%
\includegraphics[width=\columnwidth]{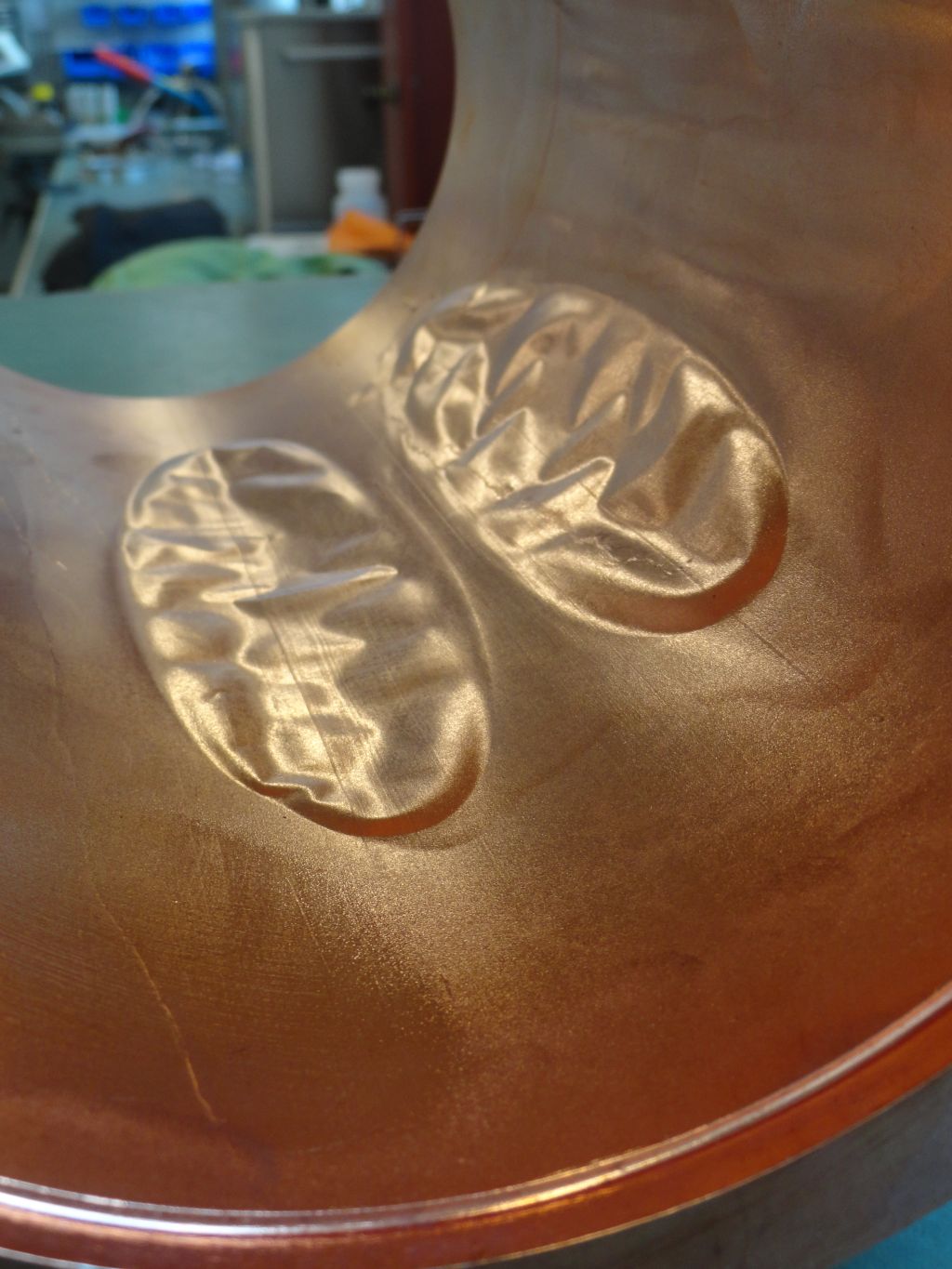}%
\caption{Photograph of the tube's internal surface after opening. Two `blisters' had developed over a combined area of $\sim \unit[150]{cm^2}$.}%
\label{fig:10}%
\end{figure}

After opening the system we observed two large `blisters' in the coating, see Fig.\,\ref{fig:10}, with an overall area of about $\unit[150]{cm^2}$. The de-laminated Cu-layer was still intact and robust; no obvious cracks or holes were visible. From the pressure evolution we had indications that some de-lamination may have happened already during the second heating cycle. Based on this we speculate that the pressure spikes discussed in connection with Fig.\,\ref{fig:9} could represent a stepwise further enlargement of the blisters.

.

\section*{Summary and Conclusion}

Without baking the vacuum properties of recipients made of copper-coated carbon reinforced plastics were comparable to recipients made of stainless steel. All tests were performed only with fluoroelastomer gaskets. Permeability and leakage rates were identical to values known for this type of sealing. A design including metallic sealing appears straight forward. After baking, out-gassing rates smaller than $\unit[10^{-8}]{mbar\!\cdot\!liter/s}$ were measured for the recipients.
No helium permeation could be measured through the CuCFRP\@.
After baking ultra-high vacuum conditions at $\unit[2\!\times\!10^{-8}]{mbar}$ were reached. It was shown that standard vacuum tubes can be easily produced from
off-the-shelf CFRP tubes and our recipients showed no measurable deformation when evacuated. Even after two years of exposure to air the vacuum performance was comparable to stainless steel.
That the coating had de-laminated over a specific area points to locally insufficient CFRP surface conditions necessary for coating which might be a direct result of the production method. The custom made sample (fig.\,\ref{fig:1b}) had a much smoother inner surface before coating. The adhesion of the copper coating might depend directly on the surface quality of the mold and the temperature during the autoclave CFRP production process. This assumption is supported by the observation that the custom made recipient, which was heated to \unit[200]{\celsius} during production, did not show any degradation during baking at higher temperatures. The precise production methods for the CFRP tube are not known to the authors, however, as for most CFRP processes a production temperatures above \unit[130]{\celsius} is unlikely. The producer of the tube, C-Tech\footnotemark[1], gives a maximum continuous temperature resistance up to \unit[115]{\celsius}, which might be too low for vacuum baking.

We noted that baking of all copper coatings yielded a largely improved surface resistance against typical degradation, as e.g.\ finger prints on untreated Cu surfaces.
While writing this article we noticed, that the Frauenhofer
Institute at Braunschweig was successful in developing copper-coatings to CFRP to produce antennas\footnote{http://www.ist.fraunhofer.de/de/pressemitteilungen/funkender-kunststoff--einzigartige-antennen-aus-cfk-und-metall.html}.

\section*{Acknowledgments}
We would like to thank P.~Fazio from Pauco Plast for his support and advice during the design and production phase of the CFRP samples. This work was in part supported by the SNF under grant \#200020-148473.
%
%

\end{document}